\documentclass[twocolumn,aps,prb]{revtex4-2}
\usepackage{ifxetex}
\ifxetex
  \usepackage{fontspec}
\else
  \usepackage{times}
\fi
\usepackage{amsmath,amsfonts,amssymb}
\usepackage{dsfont}
\usepackage{amsthm,mathrsfs}
\usepackage{soul,bm,array,graphicx,bbold,multirow}
\usepackage[normalem]{ulem}
\usepackage[makeroom]{cancel}
\usepackage[usenames,dvipsnames]{xcolor}
\usepackage[colorlinks=true,citecolor=blue,linkcolor=red]{hyperref}
\usepackage{pdfpages}
\usepackage{graphicx}     
\usepackage[font=small,labelfont=bf,labelsep=period]{caption}
\usepackage{subcaption}   
\makeatletter
\AtBeginDocument{\let\LS@rot\@undefined}
\makeatother
\newcolumntype{x}[1]{>{\centering\let\newline\\\arraybackslash\hspace{0pt}}p{#1}}

\DeclareMathAlphabet{\mathbbold}{U}{bbold}{m}{n}

\newcounter{subeqn} %

\makeatletter
\@addtoreset{subeqn}{equation}
\def\l@subsection#1#2{}
\def\l@subsubsection#1#2{}
\makeatother

\usepackage{ragged2e}
\makeatletter
\def\frontmatter@title@format{\large\bfseries\centering\parskip\z@skip}%
\def\section{%
  \@startsection{section}{1}{\z@}{0.8cm \@plus1ex \@minus .2ex}{0.5cm}%
  {\normalfont\small\bfseries\centering}%
}%
\def\@hangfrom@section#1#2#3{\@hangfrom{#1#2}\MakeTextUppercase{#3}}%
\def\@hangfroms@section#1#2{#1\MakeTextUppercase{#2}}%
\def\subsection{%
  \@startsection{subsection}{2}{\z@}{.8cm \@plus1ex \@minus .2ex}{.5cm}%
  {\normalfont\small\bfseries\centering}%
}%
\def\subsubsection{%
  \@startsection{subsubsection}{3}{\z@}{.8cm \@plus1ex \@minus .2ex}{.5cm}%
  {\normalfont\small\itshape\centering}%
}%
\makeatother
\begin{document}
\title{Non-Hermitian Quantum Mechanics I: Instantaneous Self-Energy}
\author{Lingfeng Liu$^1$}
\author{Wei-Wei Yang$^2$}
\email[Corresponding author: ]{yangww18@lzu.edu.cn}
\author{Jiangping Hu$^{2,3}$}
\email[Corresponding author: ]{jphu@iphy.ac.cn}
\author{Zhesen Yang$^{1,4}$}
\email[Corresponding author: ]{yangzs@xmu.edu.cn}
\affiliation{$^1$ Department of Physics, Xiamen University, Xiamen 361005, Fujian Province, China}
\affiliation{$^2$ Beijing National Laboratory for Condensed Matter Physics and Institute of Physics, Chinese Academy of Sciences, Beijing 100190, China}
\affiliation{$^3$ New Cornerstone Science Laboratory, Beijing, 100190, China}
\affiliation{$^4$ Asia Pacific Center for Theoretical Physics, Pohang, Republic of Korea}
\date{\today}

\begin{abstract}
Starting from the unitary evolution of a closed quantum system, we rigorously demonstrate that the projection of the global wavefunction onto an arbitrary local subsystem is governed by an exact,  time-dependent non-Hermitian Schr\"odinger equation.
Crucially, the derivation does not rely on conventional approximations such as the Born approximation, the Markov approximation, or the wide-band limit.
The central quantity is the \textit{instantaneous self-energy}, a time-dependent and generally non-Hermitian operator that encodes environmental backaction and, together with the subsystem Hamiltonian, forms the exact time-local generator of the projected dynamics.
By benchmarking the conventional non-Hermitian approximation against this exact framework, we systematically expose its limitations.
These results provide a rigorous microscopic foundation for the emergence of effective non-Hermitian dynamics in quantum systems.
\end{abstract}
\maketitle

\makeatletter
\def\l@subsection#1#2{}
\def\l@subsubsection#1#2{}
\def\l@section#1#2{%
	\vspace{1.5pt}%
	\begingroup
	\parindent \z@ 
	\rightskip \@pnumwidth
	\parfillskip -\@pnumwidth
	\leavevmode
	\advance\leftskip 2.4em\relax
	\hskip -\leftskip
	#1\nobreak\hfil \nobreak\hb@xt@\@pnumwidth{\hss #2}\par
	\endgroup}
\makeatother
\vspace{1em}

\begin{figure}[t]
	\includegraphics[width=\columnwidth]{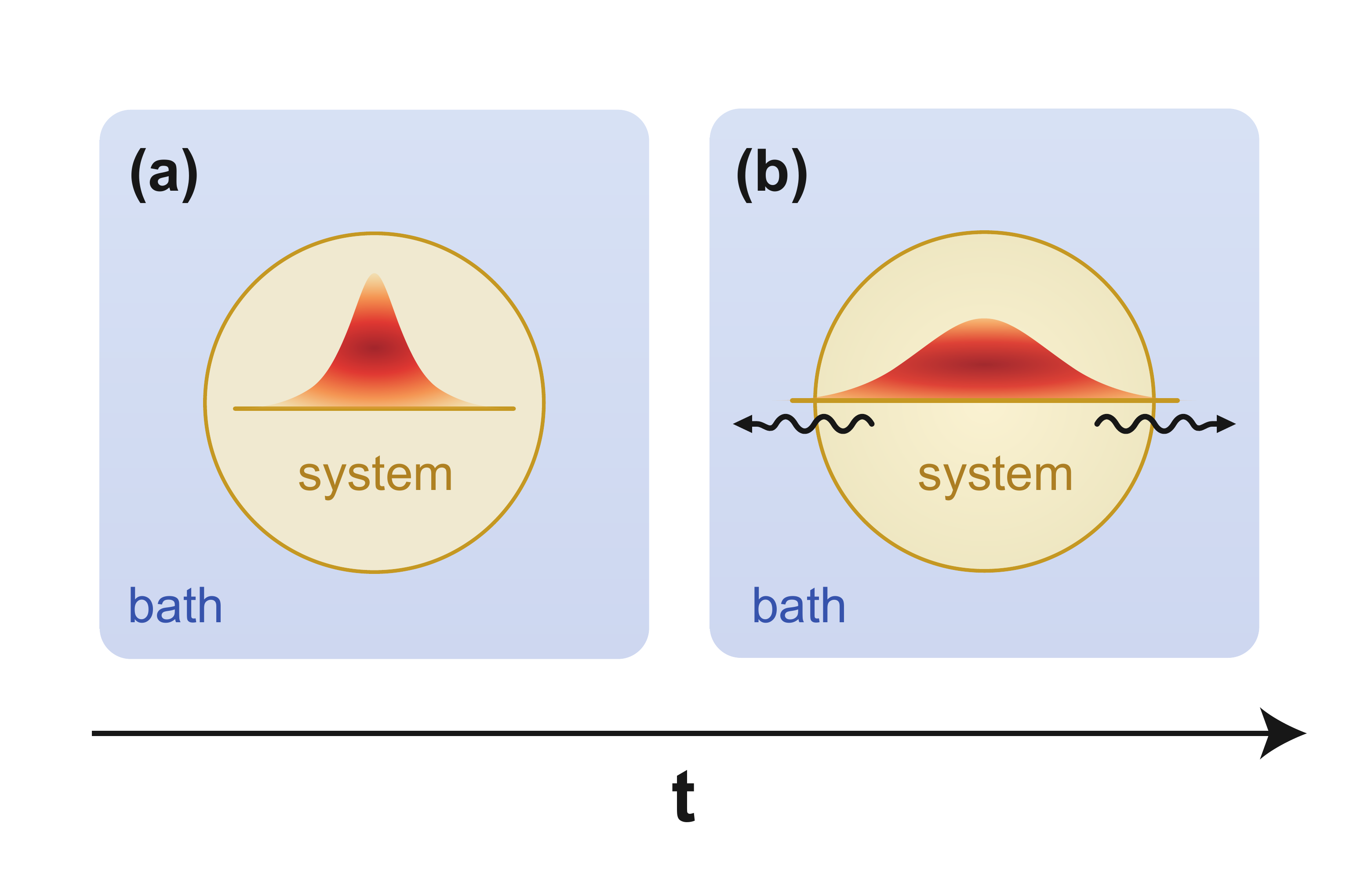}
	\caption{\justifying
		Schematic illustration of the projected non-Hermitian dynamics. The blue region represents the environment and the yellow circle the local subsystem. 
		(a) The wavefunction is initially localized within the subsystem.
		(b) Probability leaks out of the subsystem at early times, constituting the microscopic origin of effective non-Hermitian dynamics.}
	\label{F1}
\end{figure}

\section{Introduction}
In recent years, non-Hermitian physics has advanced rapidly across a wide range of classical wave systems, including photonics, acoustics, electrical circuits, and mechanical metamaterials~\cite{Ruter2010NatPhys,Zhu2014PRX,Schindler2011PRA,Brandenbourger2019NatCommun,Liu2021Research,Nature2025,Konotop2016RMP,ElGanainy2018NatPhys,Ashida2020NonHermitianP}.
These developments have revealed a wealth of characteristic non-Hermitian phenomena, notably exceptional points (EPs)~\cite{Dembowski2001PRL,Miri2019Science,Heiss_2012,Ding2022NRP,LiNN2023,NM2019,HodaeiNature2017,PhysRevLett.127.186601,Ashida2020NonHermitianP} and the non-Hermitian skin effect (NHSE)~\cite{PhysRevLett.121.086803,Kunst2018PRL,Bergholtz2021RMP,Zhang2022AdvPhysX,Kawabata2019PRX,Yang2020PRL,Zhang2020PRL,ZhangNc2022,Gohsrich_2025,PhysRevB.99.201103,PhysRevLett.121.136802,Song2019PRL,PhysRevLett.124.086801,PhysRevB.102.205118,Ashida2020NonHermitianP}.
Despite this progress, a fundamental question remains: must non-Hermitian quantum mechanics necessarily be an approximate effective theory~\cite{Bender1998PRL,Mostafazadeh2002JMP,Daley2014AdvPhys}?
More specifically, can an exact non-Hermitian Schr\"odinger equation be derived from the standard Schr\"odinger equation governing a closed quantum system, without invoking any approximation?
At first sight, the answer appears to be negative.
In standard quantum mechanics, the Hamiltonian of a closed system is required to be Hermitian, ensuring unitary time evolution and a real energy spectrum~\cite{Bender1998PRL,Mostafazadeh2002JMP,Bender2007RPP,Bender2024RMP}.
Introducing non-Hermiticity directly into the Hamiltonian therefore appears to conflict with these fundamental principles.
The key observation, however, is that Hermiticity and global probability conservation apply only to fully isolated systems~\cite{Breuer2002book}.

As illustrated in Fig.~\ref{F1}, the situation changes qualitatively once the global wavefunction is \textit{projected} onto a local subsystem~\cite{Feshbach1958,Feshbach1962,Fano1961,Rotter2009JPA,Rotter2015}.
Under this spatial projection, probability amplitude can flow out of the subsystem and into the surrounding environment.
For a finite environment, memory effects can also induce a coherent backflow of probability into the subsystem at later times~\cite{Breuer2009PRL,Breuer2016RMP,deVega2017RMP,Lambropoulos2000}.
This continuous exchange of probability between the subsystem and its environment generally renders the projected dynamics nonunitary and naturally gives rise to an effective non-Hermitian generator~\cite{Plenio1998RMP,Rotter2009JPA,Daley2014AdvPhys,Rotter2015,Ashida2020NonHermitianP}.

Although this physical picture is transparent, a rigorous microscopic derivation of the exact non-Hermitian Schr\"odinger equation is still lacking.
In practice, non-Hermitian Hamiltonians are often introduced phenomenologically by adding an imaginary potential or a complex on-site energy to the subsystem Hamiltonian~\cite{Feshbach1954PR,Makris2008PRL,Guo2009PRL,Rudner2009PRL,Moiseyev2011book,Schomerus:13,ElGanainy2018NatPhys,PhysRevA.94.022119}.
The associated parameters are then fitted to experimental data or chosen phenomenologically.
This approach, however, leaves several fundamental questions unresolved:

(i)~What physical information is lost when the exact projected dynamics is replaced by a phenomenological non-Hermitian Hamiltonian?

(ii)~Under what conditions does the phenomenological description fail qualitatively?
For example, following a quantum quench~\cite{Calabrese2006PRL,Polkovnikov2011RMP}, do the effective non-Hermitian parameters remain invariant in time, or must they be dynamically reconstructed?

(iii)~How can the phenomenological description be systematically corrected or extended when it fails?

These questions motivate the present series of papers, entitled \textit{Non-Hermitian Quantum Mechanics} (NHQM), which aims to address them systematically and rigorously.
Specifically, questions~(i) and~(ii) are addressed in the present paper (\textit{NHQM~I}) and its companion paper, \textit{NHQM~II}, while question~(iii) will be addressed in \textit{NHQM~III}.

This paper is organized as follows.
In Sec.~\ref{sec:emergence}, we present an exact and self-consistent derivation of the non-Hermitian Schr\"odinger equation directly from the unitary dynamics of a closed global system, without introducing phenomenological assumptions, perturbative expansions, or adjustable parameters.
Our framework is formulated in a first-quantized, single-particle setting with a direct-sum Hilbert space, $\mathcal{H} = \mathcal{H}_{\rm S} \oplus \mathcal{H}_{\rm B}$.
On every time interval where the projected propagator remains invertible, the projected dynamics of $|\Psi_{\rm S}(t)\rangle$ is governed by the exact time-local effective Hamiltonian $H_{\rm eff}(t) = H_{\rm S} + \Sigma_{\rm inst}(t)$ for initial states localized entirely within the subsystem.
Here, $\Sigma_{\rm inst}(t)$ is the \textit{instantaneous self-energy}, an explicitly time-dependent and generally non-Hermitian operator that encodes the environmental backaction on the subsystem~\cite{Clerk2010RMP}.
For an initially populated environment, the same generator appears together with an additional source term.

In Sec.~\ref{sec:NHA}, we review the conventional non-Hermitian approximation (NHA) and connect its weak-coupling, on-shell form to Fermi's golden rule~\cite{WeisskopfWigner1930}.
In Sec.~\ref{sec:breakdown}, we benchmark the static NHA against the exact instantaneous self-energy in the Anderson--Fano model~\cite{Anderson1961,Fano1961} using two examples: single-impurity decay and a sudden quench of the impurity potential.
The decay example illustrates the short-time buildup of environmental backaction, intermediate-time deviations from the on-shell prediction, late-time memory effects~\cite{Fonda1978,Kofman2000,Rothe2006,Longhi2006PRL}, and finite-size recurrences~\cite{Bocchieri1957PR}.
The quench example shows that the energy shift and decay rate remain continuous across the quench and then relax toward new quasisteady values.
Finally, Sec.~\ref{sec:summary} discusses the scope and implications of the framework and outlines directions for the subsequent papers in this series.
\section{Time-dependent Non-Hermitian Hamiltonians}\label{sec:emergence}
Consider a closed single-particle quantum system whose Hilbert space admits an orthogonal direct-sum decomposition $\mathcal{H} = \mathcal{H}_{\rm S} \oplus \mathcal{H}_{\rm B}$. Here, $\mathcal{H}_{\rm S}$ describes a local subsystem $\rm S$ and $\mathcal{H}_{\rm B}$ its environment $\rm B$.
Let $P_{\rm S}$ and $P_{\rm B}$ denote the corresponding projection operators. 
We define the projected Hamiltonian blocks as $H_{XY} \!\equiv\! P_X H P_Y$, where $X, Y \!\in\! \{{\rm S}, {\rm B}\}$.
With $\hbar=1$, the time-dependent Schr\"odinger equation takes the block-matrix form
\begin{equation}
	i\frac{\partial}{\partial t}
	\begin{pmatrix}
		|\Psi_{\rm S}(t)\rangle \\[2pt]
		|\Psi_{\rm B}(t)\rangle
	\end{pmatrix}
	=
	\begin{pmatrix}
		H_{\rm S} & H_{\rm SB} \\[2pt]
		H_{\rm BS} & H_{\rm B}
	\end{pmatrix}
	\begin{pmatrix}
		|\Psi_{\rm S}(t)\rangle \\[2pt]
		|\Psi_{\rm B}(t)\rangle
	\end{pmatrix}.
	\label{E1}
\end{equation}
Here, $|\Psi_{\rm S}(t)\rangle \equiv P_{\rm S}|\Psi(t)\rangle$ and $|\Psi_{\rm B}(t)\rangle \equiv P_{\rm B}|\Psi(t)\rangle$ are the projected components of the global wavefunction.
Throughout this section, the total Hamiltonian is Hermitian and time independent.

\subsection{Homogeneous Case: Theorem I}
We first consider the case in which the initial state is entirely localized in the subsystem,
\begin{equation}
	|\Psi_{\rm S}(0)\rangle = |\Phi_{\rm S}\rangle,\quad  |\Psi_{\rm B}(0)\rangle =0.
	\label{E2}
\end{equation}
Under global unitary evolution, probability amplitude can leak from the subsystem into the environment.
Let $U_{XY}(t)\equiv P_X e^{-iHt} P_Y$ denote the projected blocks of the global time-evolution operator.\\
\noindent\textit{\bf Theorem I}: Assume that the subsystem Hilbert space is finite dimensional.
Over any time interval where $U_{\rm SS}(t)$ is invertible, the projected wavefunction exactly obeys the time-local non-Hermitian Schr\"odinger equation:
\begin{equation}
	\begin{aligned}
		i\partial_t |\Psi_{\rm S}(t)\rangle
		&=
		\big[H_{\rm S}+\Sigma_{\rm inst}(t)\big]
		|\Psi_{\rm S}(t)\rangle \\[2pt]
		&\equiv
		H_{\rm eff}(t)|\Psi_{\rm S}(t)\rangle,
		\label{E3}
	\end{aligned}
\end{equation}
where the \textit{instantaneous self-energy} of the subsystem is given by
\vspace{2pt}
\begin{widetext}
\begin{equation}
	\begin{aligned}
		\Sigma_{\rm inst}(t) =
		&\bigg[
		\int_{-\infty}^{\infty} \!\! d\omega\,
		\Sigma_{\rm S}^{R}(\omega)
		G_{\rm S}^{R}(\omega)
		e^{-i\omega t}
		\bigg] 
		\times
		\bigg[
		\int_{-\infty}^{\infty} \!\! d\omega\,
		G_{\rm S}^{R}(\omega)
		e^{-i\omega t}
		\bigg]^{-1}, \quad t>0.
	\end{aligned}
	\label{E5}
\end{equation}
\end{widetext}
Here, the retarded Green's function of the subsystem is defined as
\begin{equation}
	G_{\rm S}^{R}(\omega)
	=
	\big[
	(\omega + i0^+)I_{\rm S}
	-H_{\rm S}
	-\Sigma_{\rm S}^{R}(\omega)
	\big]^{-1},
	\label{E6}
\end{equation}
and the retarded self-energy is
\begin{equation}
	\Sigma_{\rm S}^{R}(\omega)
	=
	\sum_m
	\frac{H_{\rm SB}|m\rangle\langle m|H_{\rm BS}}
	{\omega-\epsilon_m+i0^+},
	\label{E7}
\end{equation}
where $\epsilon_m$ and $|m\rangle$ are the eigenvalues and eigenstates of the environmental Hamiltonian $H_{\rm B}$.

Although the global Hamiltonian is Hermitian and time independent, the exact time-local generator $H_{\rm eff}(t)$ is generally time dependent and non-Hermitian.
The formal solution of Eq.~\eqref{E3} is
\begin{equation}
	|\Psi_{\rm S}(t)\rangle = \mathcal{T} \exp \left[ -i \int_0^t dt'\, H_{\rm eff}(t') \right] |\Psi_{\rm S}(0)\rangle,
	\label{E8}
\end{equation}
where $\mathcal{T}$ denotes the time-ordering operator.
The generator $H_{\rm eff}(t)$ is determined solely by the global Hamiltonian and the choice of subsystem projection, rather than by the particular initial state $|\Psi_{\rm S}(0)\rangle$.
It therefore serves as a universal generator for all initial states prepared entirely within the subsystem, provided that the dynamics is restricted to an interval on which $U_{\rm SS}(t)$ remains invertible.

We summarize the key steps of the proof below; a complete derivation is provided in Appendix~\ref{app:Proof1}.

\textit{Step~1.} Under the initial condition in Eq.~\eqref{E2}, the subsystem dynamics is governed by the projected propagator:
\begin{equation}
	|\Psi_{\rm S}(t)\rangle = U_{\rm SS}(t) |\Psi_{\rm S}(0)\rangle.
	\label{E9}
\end{equation}
Whenever $U_{\rm SS}(t)$ is invertible, the time-local generator is given by the right logarithmic derivative:
\begin{equation}
	H_{\rm eff}(t) = 
	i \big[\partial_t U_{\rm SS}(t)\big] U_{\rm SS}^{-1}(t).
	\label{E10}
\end{equation}
Subtracting the bare subsystem Hamiltonian defines the instantaneous self-energy,
\begin{equation}
	\Sigma_{\rm inst}(t) = 
	\left[ i\partial_t U_{\rm SS}(t) - H_{\rm S} U_{\rm SS}(t) \right] U_{\rm SS}^{-1}(t).
	\label{E11}
\end{equation}
At times when $U_{\rm SS}(t)$ loses invertibility, the universal generator becomes singular or undefined.

\textit{Step~2.} The projected propagator can be expressed in terms of the subsystem's frequency-domain retarded Green's function as
\begin{equation}
	U_{\rm SS}(t) = \frac{i}{2\pi} \int_{-\infty}^{\infty} d\omega\, G_{\rm S}^R(\omega) e^{-i\omega t},
	\quad t>0 .
	\label{E12}
\end{equation}
Using the exact resolvent identity
\begin{equation}
	(\omega + i0^+ - H_{\rm S}) G_{\rm S}^R(\omega) = I_{\rm S} + \Sigma_{\rm S}^R(\omega) G_{\rm S}^R(\omega),
	\label{E13}
\end{equation}
we can rewrite the numerator in Eq.~\eqref{E11} for $t>0$ as
\begin{equation}
	i\partial_t U_{\rm SS}(t) - H_{\rm S} U_{\rm SS}(t) = \frac{i}{2\pi} \int_{-\infty}^{\infty} d\omega\, \Sigma_{\rm S}^R(\omega) G_{\rm S}^R(\omega) e^{-i\omega t}.
	\label{E14}
\end{equation}
Substituting this result and Eq.~\eqref{E12} into Eq.~\eqref{E11} gives Eq.~\eqref{E5}.
No Born, Markov, wide-band, or perturbative approximation is invoked in this derivation.

\subsection{Inhomogeneous Case: Theorem II}
We now examine whether a nonzero initial environmental component alters the functional form of the instantaneous self-energy.
As shown below, it does not: the algebraic structure of $\Sigma_{\rm inst}(t)$ is strictly universal.
The initial environmental component enters only through an additional source term.\\
\noindent\textbf{Theorem II}: Consider a general initial state with subsystem and environmental components
\begin{equation}
	|\Psi_{\rm S}(0)\rangle = |\Phi_{\rm S}\rangle,\quad  |\Psi_{\rm B}(0)\rangle =|\Phi_{\rm B}\rangle.
	\label{E15}
\end{equation}
Over any time interval where $U_{\rm SS}(t)$ is invertible, the projected wavefunction exactly satisfies the inhomogeneous time-dependent non-Hermitian Schr\"odinger equation:
\begin{equation}
	\begin{aligned}
		i\partial_t |\Psi_{\rm S}(t)\rangle
		&=
		\left[H_{\rm S}+\Sigma_{\rm inst}(t)\right]|\Psi_{\rm S}(t)\rangle + |F(t)\rangle  \\[2pt]
		&\equiv
		H_{\rm eff}(t)|\Psi_{\rm S}(t)\rangle + |F(t)\rangle,
		\label{E16}
	\end{aligned}
\end{equation}
where the generator
\begin{equation}
	H_{\rm eff}(t) = H_{\rm S} + \Sigma_{\rm inst}(t) 
	\label{E17}
\end{equation}
coincides with the expression derived in Theorem~I and is independent of both $|\Phi_\mathrm{S}\rangle$ and $|\Phi_\mathrm{B}\rangle$.
The influence of the initial environmental component is entirely contained in the source term
\begin{equation}
	|F(t)\rangle = H_{\rm SB} S_{\rm B}(t) |\Psi_{\rm B}(0)\rangle,
	\label{E18}
\end{equation}
where $S_{\rm B}(t) \!\equiv\! \left[U^\dagger_{\rm BB}(t)\right]^{-1}$. 
Here, $U_{\rm BB}(t) \!\equiv\! P_{\rm B} e^{-i H t} P_{\rm B}$ is the forward projected propagator within $\mathcal{H}_{\rm B}$.
When the environmental component vanishes initially, namely, $|\Psi_{\rm B}(0)\rangle=0$, the source term vanishes, $|F(t)\rangle=0$, and Eq.~\eqref{E16} reduces to the homogeneous equation of Theorem~I.
The detailed proof of Theorem~II is presented in Appendix~\ref{app:Proof2}.

Theorem~II establishes that the global initial condition does not alter the effective generator $H_{\rm eff}(t)$.
Instead, the initial environmental component enters the projected dynamics through the source term $|F(t)\rangle$.
Therefore, for a fixed global Hamiltonian and a fixed subsystem projection, $H_{\rm eff}(t)$ is universal for all initial states on every interval where $U_\mathrm{SS}(t)$ remains invertible.
In the remainder of this work, we focus on the homogeneous case where $|\Psi_{\rm B}(0)\rangle=0$.

One may nevertheless ask what is gained by introducing $\Sigma_{\rm inst}(t)$, given that its construction requires the projected propagator $U_{\rm SS}(t)$ and its inverse.
The significance of this construction is twofold.
First, it establishes a rigorous microscopic foundation for non-Hermitian physics, providing a benchmark for systematically evaluating the validity of the phenomenological NHA, as discussed in Sec.~\ref{sec:NHA}.
Second, this explicit representation provides a starting point for systematic perturbative expansions, enabling the construction of effective non-Hermitian theories that faithfully capture transient dynamics beyond the NHA.
The corresponding perturbative framework will be developed in \textit{NHQM~III}.

\subsection{Self-Energy Decomposition and Physical Consequences}
To clarify the physical content of the effective Hamiltonian $H_{\rm eff}(t) = H_{\rm S} + \Sigma_{\rm inst}(t)$, we decompose the instantaneous self-energy into its Hermitian and anti-Hermitian parts,
\begin{equation}
	\Sigma_{\rm inst}(t) \equiv \Delta(t) - i \frac{\Gamma(t)}{2},
	\label{E19}
\end{equation}
where
\begin{equation}
	\begin{aligned}
		\Delta(t) &\equiv \frac{1}{2}\left[\Sigma_{\rm inst}(t) + \Sigma_{\rm inst}^\dagger(t)\right], \\[2pt]
		\Gamma(t) &\equiv i\left[\Sigma_{\rm inst}(t) - \Sigma_{\rm inst}^\dagger(t)\right].
	\end{aligned}
	\label{E20}
\end{equation}
Both $\Delta(t)$ and $\Gamma(t)$ are Hermitian.
The matrix $\Delta(t)$ describes the instantaneous environment-induced energy renormalization.
The matrix $\Gamma(t)$ governs the instantaneous probability exchange between the subsystem and the environment.
With this decomposition, the effective Schr\"odinger equation takes the form
\begin{equation}
	i\partial_t |\Psi_{\rm S}(t)\rangle = \left[H_{\rm S} + \Delta(t) - i\frac{\Gamma(t)}{2}\right] |\Psi_{\rm S}(t)\rangle.
	\label{E21}
\end{equation}
The survival probability in the subsystem is
\begin{equation}
	P_{\rm S}(t) \equiv \langle\Psi_{\rm S}(t)|\Psi_{\rm S}(t)\rangle.
	\label{E22}
\end{equation}
Differentiating $P_{\rm S}(t)$ and using Eq.~\eqref{E21} gives
\begin{equation}
	\partial_t P_{\rm S}(t) = -\langle\Psi_{\rm S}(t)| \Gamma(t) |\Psi_{\rm S}(t)\rangle.
	\label{E23}
\end{equation}
The expectation value of $\Gamma(t)$ therefore determines the instantaneous net probability flux across the subsystem boundary.
A positive expectation value corresponds to probability loss from the subsystem, whereas a negative value signals coherent probability backflow from the environment.

The decomposition also provides a direct characterization of the normality of the effective Hamiltonian
\begin{equation}
	\left[ H_{\text{eff}}(t), H_{\rm eff}^{\dagger}(t) \right]
	= i\left[ H_{\rm S}+\Delta(t), \Gamma(t) \right].
\end{equation}
The noncommutativity of the energy matrix $H_{\rm S}+\Delta(t)$ and the probability-exchange matrix $\Gamma(t)$ is therefore precisely equivalent to the nonnormality of $H_{\text{eff}}(t)$, as shown in Appendix~\ref{app:Proof5}.
We finally note that this non-commutativity is the foundational condition for all genuinely non-Hermitian phenomena: non-reciprocal hopping, the non-Hermitian skin effect, and EPs.

\subsection{Spectral Representation and Numerical Implementation}
We next formulate an exact time-domain procedure for evaluating $\Sigma_{\rm inst}(t)$ in a finite-dimensional global system.

We first diagonalize the global Hamiltonian $H$ to obtain its eigenvalues $E_j$ and eigenstates $|\Phi_j\rangle$:
\begin{equation}
	H |\Phi_j\rangle = E_j |\Phi_j\rangle.
	\label{E25}
\end{equation}
The projected subsystem propagator $U_{\rm SS}(t)$ can then be written as
\begin{equation}
	U_{\rm SS}(t) = \sum_j e^{-i E_j t} P_{\rm S} |\Phi_j\rangle\langle\Phi_j| P_{\rm S}.
	\label{E26}
\end{equation}
Differentiating with respect to time gives
\begin{equation}
	i\partial_t U_{\rm SS}(t) = \sum_j E_j e^{-i E_j t} P_{\rm S} |\Phi_j\rangle\langle\Phi_j| P_{\rm S}.
	\label{E27}
\end{equation}
Substituting these expressions into Eq.~\eqref{E11} yields
\begin{equation}
	\begin{aligned}
		\Sigma_{\rm inst}(t) 
		&= \left( \sum_j E_j e^{-i E_j t} P_{\rm S} |\Phi_j\rangle\langle\Phi_j| P_{\rm S} \right) \\[2pt]
		&\quad \times \left( \sum_k e^{-i E_k t} P_{\rm S} |\Phi_k\rangle\langle\Phi_k| P_{\rm S} \right)^{-1} - H_{\rm S}.
	\end{aligned}
	\label{E28}
\end{equation}
This expression is exact on every time interval over which $U_{\rm SS}(t)$ remains invertible.

The spectral construction has two principal advantages.
First, it is free of artificial broadening parameters and retains the exact discrete spectrum of the finite global Hamiltonian.
Second, it captures the full projected dynamics, including Poincar\'e recurrences.
All subsequent finite-size numerical calculations use this spectral construction.

\section{Non-Hermitian Approximation}\label{sec:NHA}
We now examine how the exact time-local framework connects to the conventional NHA used in open-system and condensed-matter physics.
We first review the Feshbach projection formalism (Sec.~\ref{sec:Feshbach}).
We then show how the NHA is recovered as a static limit of our exact theory (Sec.~\ref{sec:NHA_consistency}) and relate its weak-coupling, on-shell form to Fermi's golden rule (Sec.~\ref{sec:FGR}).

\subsection{Review of the Feshbach Projection Formalism}\label{sec:Feshbach}
The Feshbach formalism eliminates environmental degrees of freedom at the level of the stationary Schrödinger equation.
Let $P_{\rm S}$ and $P_{\rm B} \equiv I - P_{\rm S}$ denote the projection operators onto the subsystem and the environment, respectively.
Applying $P_{\rm S}$ and $P_{\rm B}$ to $(E - H) |\Psi(E)\rangle = 0$ yields
\begin{equation}
	\begin{aligned}
		(E I_{\rm S} - H_{\rm S}) P_{\rm S}|\Psi(E)\rangle &= H_{\rm SB} P_{\rm B}|\Psi(E)\rangle, \\[1ex]
		(E I_{\rm B} - H_{\rm B}) P_{\rm B}|\Psi(E)\rangle &= H_{\rm BS} P_{\rm S}|\Psi(E)\rangle,
	\end{aligned}
	\label{E29}
\end{equation}
where the Hamiltonian blocks are
\begin{equation}
	\begin{aligned}
		H_{\rm S} &= P_{\rm S}HP_{\rm S}, \quad H_{\rm B} = P_{\rm B}HP_{\rm B}, \\[2pt]
		H_{\rm SB} &= P_{\rm S}HP_{\rm B}, \quad H_{\rm BS} = P_{\rm B}HP_{\rm S}.
	\end{aligned}
\end{equation}
For $E$ in the resolvent set of $H_{\rm B}$, solving the second equation for the environmental component gives
\begin{equation}
	P_{\rm B} \left| \Psi(E) \right\rangle =
	\left( E I_{\rm B} - H_{\rm B} \right)^{-1} H_{\rm BS} P_{\rm S} \left| \Psi(E) \right\rangle.
	\label{E31}
\end{equation}
Substituting this result into the first equation yields the exact projected equation
\begin{equation}
	\left[ E I_{\rm S} - H_{\rm S} - H_{\rm SB}(E I_{\rm B} - H_{\rm B})^{-1} H_{\rm BS} \right] 
	P_{\rm S}|\Psi(E)\rangle = 0.
	\label{E32}
\end{equation}
The effect of the environment is thus encoded in an energy-dependent operator acting within the subsystem Hilbert space.
The retarded prescription $E \to E + i0^+$ defines the self-energy
\begin{equation}
	\Sigma_{\rm S}^R(E) = 
	H_{\rm SB} 
	[(E+i0^+)I_{\rm B} - H_{\rm B}]^{-1} 
	H_{\rm BS}.
	\label{E33}
\end{equation}
The corresponding projected retarded Green's function is
\begin{equation}
	G_{\rm S}^R(E) = \left[ (E+i0^+)I_{\rm S} - H_{\rm S} - \Sigma_{\rm S}^R(E) \right]^{-1}.
\end{equation}
The infinitesimal positive imaginary part ensures causality.
This frequency-domain Feshbach representation is exact.
The time-domain retarded Green's function satisfies
\begin{equation}
	G_{\rm S}^R(t) = -i \theta(t) U_{\rm SS}(t).
\end{equation}
Inverse Fourier transformation therefore gives the projected propagator,
\begin{equation}
	U_{\rm SS}(t) = \frac{i}{2\pi} \int_{-\infty}^{\infty} dE\, G_{\rm S}^R(E) e^{-i E t}, \quad t>0.
	\label{E36}
\end{equation}
For an initially unpopulated environment, the subsystem wavefunction is then
\begin{equation}
	|\Psi_{\rm S}(t)\rangle = U_{\rm SS}(t) |\Psi_{\rm S}(0)\rangle.
	\label{E37}
\end{equation}

Although the Feshbach representation is exact, obtaining explicit expressions for the real-time dynamics presents several analytical challenges:

(i) For structured environments, such as multidimensional lattices, disordered systems, and complex molecular reservoirs, the self-energy $\Sigma_{\rm S}^R(E)$ may not admit a closed-form expression.

(ii) Even when a closed-form expression for the self-energy is available, spectral features such as cusps or van Hove singularities can make the Fourier integral analytically intractable.

(iii) For a subsystem with $N_{\rm S}>1$ levels, the Green's function is the $N_{\rm S}\times N_{\rm S}$ energy-dependent matrix $[(E+i0^+)I_{\rm S} - H_{\rm S} - \Sigma_{\rm S}^R(E)]^{-1}$.
As $N_{\rm S}$ increases, evaluating this matrix inverse and the subsequent frequency integral in closed form becomes increasingly cumbersome.
These analytical difficulties often necessitate numerical evaluation of the projected resolvent or propagator.

A central difficulty is that the self-energy $\Sigma_{\rm S}^R(E)$ appears in the matrix denominator of the Green's function.
Its nontrivial analytic structure can prevent the Fourier integral in Eq.~\eqref{E36} from being evaluated in closed form.
Our time-domain formulation offers an alternative route to the projected dynamics.
In \textit{NHQM~III}, we will develop a time-domain perturbation theory that avoids complex analytic continuation in the weak-coupling regime.
We further apply this framework to investigate weak-coupling non-Markovian systems in \textit{NHQM~V}.

\subsection{The Non-Hermitian Approximation and Its Self-Consistency}\label{sec:NHA_consistency}
An analytically tractable approximation to the projected dynamics is obtained by replacing the energy-dependent retarded self-energy with a constant matrix,
\begin{equation}
	\Sigma_{\rm S}^R(E) \approx \Sigma_0.
	\label{E38}
\end{equation}
Within this static approximation, the projected Green's function simplifies to
\begin{equation}
	\begin{aligned}
		G_{\rm S}^R(E) &= \left[(E+i0^+)I_{\rm S} - H_{\rm S} - \Sigma_{\rm S}^R(E)\right]^{-1} \\
		&\approx \left[E - H_{\rm S} - \Sigma_0\right]^{-1} \\
		&\equiv G^R_0(E).
	\end{aligned}
\end{equation}
We define the static non-Hermitian Hamiltonian
\begin{equation}
	H_{\rm nH} = H_{\rm S} + \Sigma_0,
\end{equation}
so that
\begin{equation}
	G_0^R(E) = \left[(E+i0^+)I_{\rm S} - H_{\rm nH}\right]^{-1}.
\end{equation}
Substituting $G^R_0(E)$ into Eq.~\eqref{E36} and evaluating the integral using Cauchy's residue theorem for $t>0$ gives
\begin{equation}
	\begin{aligned}
		U_{\rm SS}(t) &\approx e^{-i\left(H_{\rm S}+\Sigma_0\right)t}.
	\end{aligned}
	\label{E42}
\end{equation}
The projected wavefunction therefore obeys the static non-Hermitian Schrödinger equation
\begin{equation}
	i\partial_t |\Psi_{\rm S}(t)\rangle = H_{\rm nH} |\Psi_{\rm S}(t)\rangle.
\end{equation}

We next verify that the same static approximation reduces the instantaneous self-energy to $\Sigma_0$.
Substituting $\Sigma_{\rm S}^R(E) \approx \Sigma_0$ and $G_{\rm S}^R(E) \approx G^R_0(E)$ 
into Eq.~\eqref{E5}, we obtain
\begin{widetext}
	\begin{equation}
		\begin{aligned}
			\Sigma_{\rm inst}(t) &= \left( \frac{i}{2\pi} \int_{-\infty}^{\infty} dE\, e^{-i E t} \Sigma_{\rm S}^R(E) G_{\rm S}^R(E) \right) \left( \frac{i}{2\pi} \int_{-\infty}^{\infty} dE\, e^{-i E t} G_{\rm S}^R(E) \right)^{-1} \\
			&\approx \Sigma_0 \left( \frac{i}{2\pi} \int_{-\infty}^{\infty} dE\, e^{-i E t} G_0^R(E) \right) \left( \frac{i}{2\pi} \int_{-\infty}^{\infty} dE\, e^{-i E t} G_0^R(E) \right)^{-1} \\
			&= \Sigma_0.
		\end{aligned}
		\label{E44}
	\end{equation}
\end{widetext}
Thus, neglecting the energy dependence of the retarded self-energy consistently recovers the static NHA within our time-local framework.

\subsection{Microscopic Route to the NHA: Fermi's Golden Rule and the On-Shell Approximation}\label{sec:FGR}
The static approximation leaves one question open: how is the constant self-energy $\Sigma_0$ determined microscopically?
For a single discrete level weakly coupled to a smooth environmental continuum, the on-shell approximation underlying Fermi's golden rule (FGR) provides a standard prescription.
Consider a subsystem consisting of a single state $|0\rangle$ with bare energy $V_0$, coupled to environmental eigenstates $\{|\phi_m\rangle\}$ with energies $\{E_m\}$.
The total Hamiltonian is
\begin{equation}
	\begin{aligned}
		H = \;& V_0 |0\rangle\langle 0| + \sum_m E_m |\phi_m\rangle\langle\phi_m| \\
		& + \sum_m \left( t_m |0\rangle\langle\phi_m| + t_m^* |\phi_m\rangle\langle 0| \right),
	\end{aligned}
	\label{E45}
\end{equation}
where $t_m \equiv \langle 0 | H_{\rm SB} | \phi_m \rangle$ is the coupling amplitude between the subsystem level and environmental mode $m$.
In the continuum and weak-coupling limits, FGR gives the leading-order transition rate out of the initially occupied state $|0\rangle$,
\begin{equation}
	\begin{aligned}
		\Gamma_{\rm FGR} &= 2\pi \sum_m |\langle \phi_m | H_{\rm BS} | 0 \rangle|^2 \delta(E_m - V_0) \\
		&= 2\pi \sum_m |t_m|^2 \delta(E_m - V_0).
	\end{aligned}
	\label{E46}
\end{equation}
For a single-level subsystem, the retarded self-energy is the scalar
\begin{equation}
	\Sigma_{\rm S}^R(E) = \left\langle 0 \left| H_{\rm SB} \frac{1}{E - H_{\rm B} + i0^+} H_{\rm BS} \right| 0 \right\rangle,
\end{equation}
and the Sokhotski--Plemelj identity,
\begin{equation}
	\frac{1}{x + i0^+} = \mathcal{P}\frac{1}{x} - i\pi \delta(x),
\end{equation}
allows us to write
\begin{equation}
	\Sigma_{\rm S}^R(E) = \Delta(E) - i\frac{\Gamma(E)}{2},
\end{equation}
where the dispersive energy shift $\Delta(E)$ and the decay rate $\Gamma(E)$ are given by
\begin{equation}
	\begin{aligned}
		\Delta(E) &= \mathcal{P} \left\langle 0 \left| H_{\rm SB} \frac{1}{E - H_{\rm B}} H_{\rm BS} \right| 0 \right\rangle \\[2pt]
		&= \mathcal{P} \sum_m \frac{|t_m|^2}{E - E_m}, \\[2pt]
		\Gamma(E) &= 2\pi \left\langle 0 \left| H_{\rm SB} \delta(E - H_{\rm B}) H_{\rm BS} \right| 0 \right\rangle \\[2pt]
		&= 2\pi \sum_m |t_m|^2 \delta(E - E_m).
	\end{aligned}
	\label{E50}
\end{equation}
Evaluating $\Gamma(E)$ on shell at the bare subsystem energy $E = V_0$ gives
\begin{equation}
	\Gamma(V_0) = 2\pi \sum_m |t_m|^2 \delta(E_m - V_0) \equiv \Gamma_{\rm FGR},
	\label{E51}
\end{equation}
and hence
\begin{equation}
	\Gamma_{\rm FGR} = -2 \operatorname{Im}[\Sigma_{\rm S}^R(V_0)].
\end{equation}
Evaluating the real part on shell gives the energy shift
\begin{equation}
	\Delta_{\rm FGR} \equiv \operatorname{Re}[\Sigma_{\rm S}^R(V_0)] = \mathcal{P} \sum_m \frac{|t_m|^2}{V_0 - E_m},
\end{equation}
which is the standard second-order Lamb shift of the subsystem level.

The on-shell approximation therefore fixes the constant self-energy as
\begin{equation}
	\Sigma_0 =
	\Sigma_{\rm S}^R(V_0) \equiv \Delta_{\rm FGR} - i\frac{\Gamma_{\rm FGR}}{2}.
	\label{E54}
\end{equation}
Replacing $\Sigma_{\rm S}^R(E)$ with its on-shell value in Eq.~\eqref{E54} gives the static non-Hermitian Hamiltonian
\begin{equation}
	H_{\rm nH} = V_0 + \Delta_{\rm FGR} - i\frac{\Gamma_{\rm FGR}}{2}.
\end{equation}
The resulting propagator is
\begin{equation}
	U_{\rm SS}^{\rm NHA}(t) =  e^{-i\left(V_0+\Delta_{\rm FGR}\right)t} e^{-\Gamma_{\rm FGR}t/2}.
\end{equation}
The associated survival probability is therefore
\begin{equation}
	P_{\rm S}^{\rm NHA}(t) = \left|U_{\rm SS}^{\rm NHA}(t)\right|^2 = e^{-\Gamma_{\rm FGR} t}.
\end{equation}
Having established both the static construction and its weak-coupling on-shell interpretation, a fundamental question naturally arises: which physical processes in the exact dynamics are neglected by the NHA?

\section{Examples and Applications}\label{sec:breakdown}
We now assess the accuracy of the static NHA by comparing its constant self-energy $\Sigma_0$
with the exact instantaneous self-energy $\Sigma_{\rm inst}(t)$ for single-impurity decay and a sudden quantum quench.
These examples identify the dynamical processes omitted by the NHA.

\subsection{Example I: Short-Time Dynamics}\label{sec:diff1}
We begin by examining the simplest single-impurity setting to illustrate how the static NHA fails at short times. Consider a discrete level $|0\rangle$ coupled to a 1D tight-binding environment (the Anderson--Fano model), governed by the Hamiltonian:
\begin{equation}
	\begin{aligned}
		H &= V_0 |0\rangle\langle0| + t_c\sum_{j=1}^{N-1} \left( |j\rangle\langle j+1| + \text{H.c.} \right) \\
		&\quad + \lambda \left( |0\rangle\langle1| + \text{H.c.} \right).
	\end{aligned}
	\label{E58}
\end{equation}
The parameters $V_0$, $t_c$, and $\lambda$ denote the impurity energy, the hopping amplitude within the environment, and the impurity--environment coupling, respectively.
The total Hilbert-space dimension is therefore $N+1$.
We take the initial state to be localized on the impurity,
\begin{equation}
	|\Psi_{\rm S}(0)\rangle = |0\rangle,\quad 
	|\Psi_{\rm B}(0)\rangle = 0.
\end{equation}
The NHA assigns a nonzero constant decay rate $\Gamma_{\rm FGR}$ at $t = 0^+$, whereas the exact dynamics has a vanishing initial decay rate.

To characterize the initial buildup of the decay rate, we expand the exact instantaneous self-energy through order $t^2$ in the short-time limit.
Starting from
\begin{equation}
	\Sigma_{\rm inst}(t) = i \left[\partial_t U_{\rm SS}(t)\right] U_{\rm SS}^{-1}(t) - H_{\rm S},
	\label{E60}
\end{equation}
we require $U_{\rm SS}(t)$ through order $t^3$
and $U_{\rm SS}^{-1}(t)$ through order $t^2$.
The projected propagator has the expansion
\begin{equation}
	U_{\rm SS}(t) = P_{\rm S} e^{-i H t} P_{\rm S} = \sum_{n=0}^{\infty} \frac{(-it)^n}{n!} 
	P_{\rm S} H^n P_{\rm S}.
\end{equation}
The first four projected powers of the global Hamiltonian are
\begin{equation}
	\begin{aligned}
		n=0:\quad P_{\rm S} H^0 P_{\rm S} &= I_{\rm S}, \\
		n=1:\quad P_{\rm S} H^1 P_{\rm S} &= H_{\rm S}, \\
		n=2:\quad P_{\rm S} H^2 P_{\rm S} &= 
		H_{\rm S}^2 + H_{\rm SB} H_{\rm BS} \equiv  H_{\rm S}^2 + M_2, \\
		n=3:\quad P_{\rm S} H^3 P_{\rm S} &= 
		H_{\rm S}^3 + H_{\rm S} H_{\rm SB} H_{\rm BS} + H_{\rm SB} H_{\rm BS} H_{\rm S} \\
		&+ H_{\rm SB} H_{\rm B} H_{\rm BS} \\
		&\equiv H_{\rm S}^3 + \{H_{\rm S}, M_2\} + M_3.
	\end{aligned}
\end{equation}
Here, $M_2 \equiv H_{\rm SB} H_{\rm BS}$ describes a round trip from the subsystem to the environment and back, while $M_3 \equiv H_{\rm SB} H_{\rm B} H_{\rm BS}$ includes propagation within the environment between the two coupling events.
The notation $\{A, B\} \equiv AB + BA$ denotes the anticommutator.
The projected propagator is therefore
\begin{equation}
	\begin{aligned}
		U_{\rm SS}(t) &= I_{\rm S} - i H_{\rm S} t - \frac{t^2}{2} (H_{\rm S}^2 + M_2) \\
		&\quad + \frac{i t^3}{6} (H_{\rm S}^3 + \{H_{\rm S}, M_2\} + M_3) + \mathcal{O}(t^4).
	\end{aligned}
	\label{E63}
\end{equation}
Differentiating this expression gives
\begin{equation}
	\begin{aligned}
		i\partial_t U_{\rm SS}(t) &= H_{\rm S} - it\,(H_{\rm S}^2 + M_2) \\
		&\quad - \frac{t^2}{2}(H_{\rm S}^3 + \{H_{\rm S}, M_2\} + M_3) + \mathcal{O}(t^3).
	\end{aligned}
	\label{E64}
\end{equation}
Using the matrix expansion $(I_{\rm S} + Y)^{-1} = I_{\rm S} - Y + Y^2 + \mathcal{O}(t^3)$, with $Y=U_{\rm SS}(t)-I_{\rm S}=\mathcal{O}(t)$, we obtain
\begin{equation}
		U_{\rm SS}^{-1}(t) = I_{\rm S} + i t H_{\rm S} + \frac{t^2}{2}(M_2 - H_{\rm S}^2) + \mathcal{O}(t^3).
	\label{E65}
\end{equation}
Multiplying Eq.~\eqref{E64} on the right by the inverse in Eq.~\eqref{E65} and subtracting $H_{\rm S}$ gives
\begin{equation}
	\begin{aligned}
		\Sigma_{\rm inst}(t) &= i[\partial_t U_{\rm SS}(t)]\, U_{\rm SS}^{-1}(t) - H_{\rm S} \\
		&= -it\, M_2 + \frac{t^2}{2}(M_2 H_{\rm S} - M_3) + \mathcal{O}(t^3).
	\end{aligned}
	\label{E66}
\end{equation}
Eq.~\eqref{E66} reveals a clear hierarchy of timescales. 
The leading imaginary part, which dictates the instantaneous probability exchange rate, grows linearly with time:
\begin{equation}
	\operatorname{Im}[\Sigma_{\rm inst}(t)] \approx -t\, M_2.
\end{equation}
This linear growth is the signature of the quantum Zeno regime. 
Because the instantaneous decay rate vanishes at $t=0$, the survival probability initially decays quadratically as $P(t) \approx 1 - t^2 \langle M_2 \rangle$ rather than exponentially. Static Markovian approximations miss this fundamental feature. 
The leading real part, corresponding to the dynamical energy shift, appears only at quadratic order:
\begin{equation}
	\operatorname{Re}[\Sigma_{\rm inst}(t)] \approx \frac{t^2}{4}\left(\{M_2, H_{\rm S}\} - 2M_3\right)
\end{equation}
This indicates that dynamical energy renormalization builds up more slowly than dissipation at the onset of evolution.

In summary, the static NHA overlooks these early-time transient dynamics by imposing a constant decay rate starting abruptly at $t=0$. The exact time-local construction instead captures both the vanishing initial rate and the subsequent buildup of environmental backaction.

\begin{figure*}[t]
	\centering
	\includegraphics[width=0.9\textwidth]{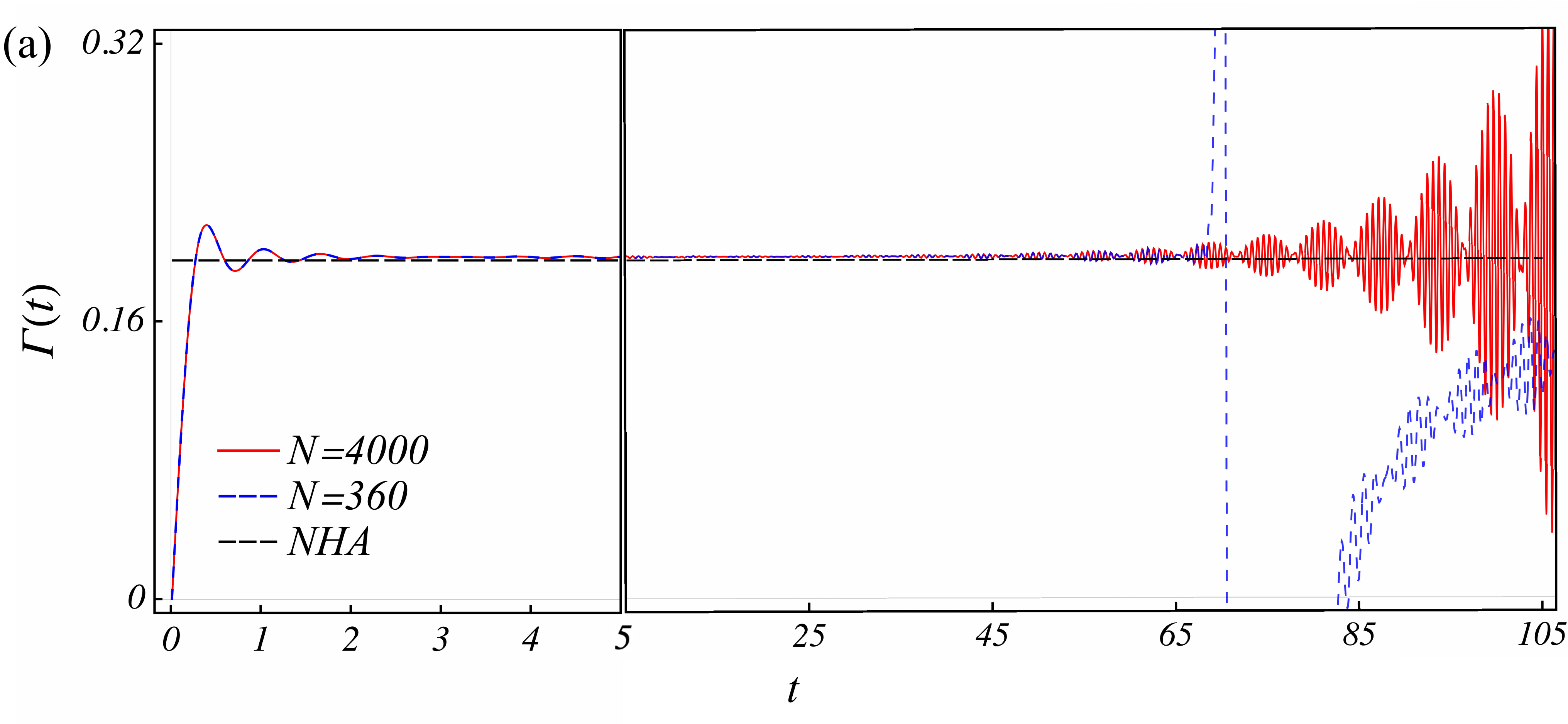}
	\caption{\justifying2
		Benchmark of the NHA in the Anderson--Fano model. 
		The decay rate $\Gamma(t)=-2\operatorname{Im}[\Sigma_{\rm inst}(t)]$ is plotted for a large environment ($N=4000$, solid red line) and a small environment ($N=360$, dashed blue line) alongside the constant on-shell FGR value $\Gamma_{\rm FGR}$ (dashed black line) for model parameters $t_c=5$, $V_0=0.5$, and $\lambda=0.7$.
		The exact dynamics exhibits an initial buildup, an intermediate-time plateau that deviates from the FGR prediction, and persistent late-time non-Markovian oscillations.
		For the smaller environment, a pronounced recurrence spike interrupts the dynamics when the wave packet returns to the impurity after reflection from the boundary, at $T_{\rm rec} \approx 72$.}
	\label{F2}
\end{figure*}

\subsubsection{Numerical Demonstration}
We test these analytical results numerically using the Anderson--Fano model in Eq.~\eqref{E58}.
We use an environment with $N=4000$ sites, a hopping amplitude $t_c=5$, an impurity energy $V_0=0.5$, and a coupling strength $\lambda=0.7$.
For an impurity energy inside the environmental continuum, $|V_0|<2t_c$, the on-shell self-energy is (see Appendix~\ref{app:FGR} for the derivation)
\begin{equation}
	\begin{aligned}
		\Sigma_{\rm S}^R(V_0) &= \frac{\lambda^2}{2t_c^2} \left( V_0 - i\sqrt{4t_c^2 - V_0^2} \right) \\
		&= \Delta_{\rm FGR} - i\frac{\Gamma_{\rm FGR}}{2}.
	\end{aligned}
\end{equation}
Fig.~\ref{F2} compares the resulting static decay rate with the time-dependent rate obtained from $\Sigma_{\rm inst}(t)$.
The comparison reveals three regimes in which the NHA misses features of the exact dynamics:

(i) \textit{Early-time breakdown:} To leading order in time, Eq.~\eqref{E66} gives $\Gamma(t) = -2\operatorname{Im}[\Sigma_{\rm inst}(t)] = 2tM_2$.
The decay rate starts from zero at $t=0$ and increases gradually, reflecting the short-time Zeno suppression absent from the static NHA.

(ii) \textit{Intermediate-time discrepancy:} After the initial transient, the exact decay rate approaches a quasisteady plateau near the on-shell FGR value $\Gamma_{\rm FGR}$, with a small offset and residual non-Markovian oscillations.
These deviations reflect off-shell energy shifts and higher-order system--bath hybridization effects omitted by the on-shell approximation.
Such corrections persist even in the weak-coupling regime.

(iii) \textit{Late-time breakdown:} The NHA also fails to capture the long-time asymptotic dynamics.
For the large environment ($N=4000$, solid red line in Fig.~\ref{F2}), the exact decay rate exhibits persistent late-time oscillations.
These oscillations reflect non-Markovian memory associated with the band-edge van Hove singularities of the tight-binding environment, which will be analyzed in detail in \textit{NHQM~II}.

To distinguish band-edge dynamics from boundary reflections, Fig.~\ref{F2} also shows the result for a shorter environment with $N=360$ sites (dashed blue line).
In this case, the recurrence time $T_{\rm rec} \approx 2N/v_g \approx 72$ (with bath group velocity $v_g = 2t_c = 10$) is shorter than the time needed for pronounced non-Markovian oscillations to develop.
Boundary reflections therefore alter the dynamics before these oscillations become prominent.
The recurrences produce periodic transient spikes, followed by a return to the Markovian plateau.
This behavior distinguishes finite-size recurrences from the persistent oscillations associated with the continuous band.

The static NHA thus misses the initial Zeno suppression, intermediate-time off-shell renormalization, and late-time non-Markovian oscillations.
A detailed analysis of these dynamical regimes and their dependence on model parameters will be presented in \textit{NHQM~II}.

\begin{figure*}[t]
	\centering
	\includegraphics[width=0.9\linewidth]{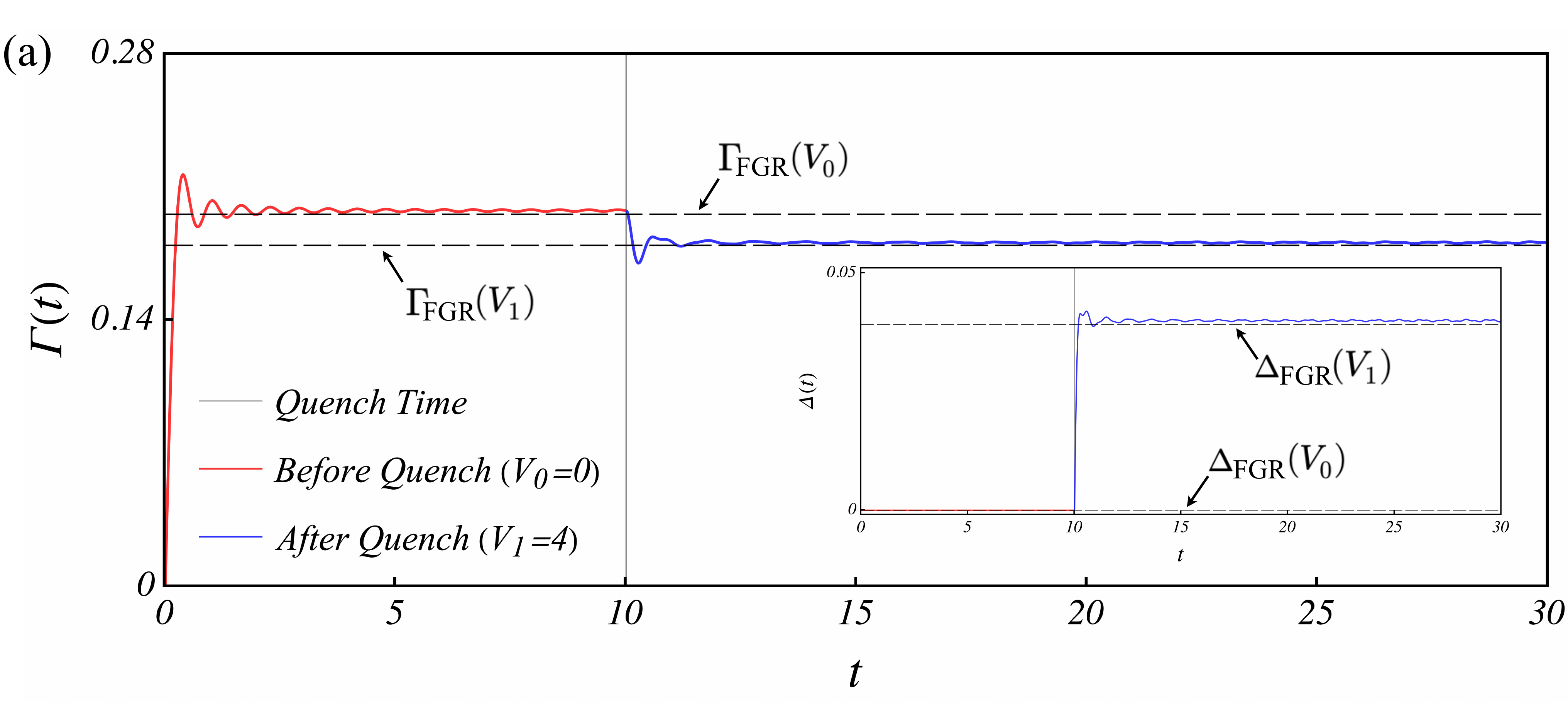}
	\caption{\justifying
		Quench-induced reconstruction of the instantaneous self-energy in the Anderson--Fano model. The impurity potential is changed from $V_0=0$ to $V_1=4$ at $t_1=10$, with $t_c=5$ and $\lambda=0.7$. 
		Both $\Gamma(t)=-2\,{\rm Im}\,\Sigma_{\rm inst}(t)$ and $\Delta(t)={\rm Re}\,\Sigma_{\rm inst}(t)$ remain continuous across the quench and then undergo transient relaxation toward their respective post-quench plateaus.
		The parenthetical labels $(V_0)$ and $(V_1)$ attached to $\Gamma_{\rm FGR}$ and $\Delta_{\rm FGR}$ distinguish their values at the pre- and post-quench impurity potentials, respectively.}
	\label{F9}
\end{figure*}

\subsection{Example II: Quench Dynamics and Dynamical Reconstruction of the Instantaneous Self-Energy}\label{sec:diff5}
A further limitation of the static NHA arises in explicitly time-dependent protocols, such as sudden quantum quenches.
Phenomenological non-Hermitian models often treat the dissipative environment as a static background.
In this description, external driving modifies only the bare subsystem Hamiltonian, while the environment-induced decay rates and energy shifts remain fixed.

To test this assumption, we apply a sudden quench to the impurity potential in the Anderson--Fano model and follow the response using our exact framework.
The global time-dependent Hamiltonian is
\begin{equation}
	\begin{aligned}
		H(t) &= V(t)|0\rangle\langle0| + \lambda \left(|0\rangle\langle1|+{\rm H.c.}\right) \\
		&\quad + t_c\sum_{j=1}^{N-1} \left(|j\rangle\langle j+1|+{\rm H.c.}\right),
	\end{aligned}
\end{equation}
with the time-dependent potential
\begin{equation}
	V(t) = V_0\Theta(t_1-t) + V_1\Theta(t-t_1).
\end{equation}
Here, $\Theta$ denotes the Heaviside step function, and $t_1$ is the quench time.
We choose $V_0=0,V_1=4,t_1=10,t_c=5$, and $\lambda=0.7$.
The quench occurs at $t_1\!=\!10$, during the quasisteady plateau and before pronounced band-edge oscillations develop.
This timing separates the response to the quench from both the initial buildup and the late-time band-edge dynamics.

Fig.~\ref{F9} shows the time evolution of the decay rate and the energy shift, revealing three features missed by the static NHA:

\textit{(i) Dynamical reconstruction of dissipation:}
The decay rate $\Gamma(t)$ evolves from its pre-quench plateau toward a new post-quench value.
This evolution shows that the environment-induced dissipation responds dynamically to the perturbation.

\textit{(ii) Driving-induced Lamb shift:} 
The Lamb shift $\Delta(t)$ also evolves toward a new quasisteady plateau, showing that the quench dynamically modifies the level renormalization.

\textit{(iii) Continuity across the quench:}
Neither $\Gamma(t)$ nor $\Delta(t)$ exhibits a discontinuous jump at $t_1$. 
Both quantities subsequently undergo oscillatory relaxation toward their post-quench plateaus.

These results show that keeping the self-energy fixed misses the dynamical response of the environment to the quench.
The perturbation modifies both the Hermitian energy shift and the anti-Hermitian dissipation term in $H_{\rm eff}(t)$, thereby changing the instantaneous spectrum and subsequent relaxation.
Our exact time-dependent framework captures this nonequilibrium response.

\subsection{Summary}
The comparison with the exact instantaneous self-energy identifies two main limitations of the static NHA:
\begin{enumerate}
	\renewcommand{\labelenumi}{(\roman{enumi})}
	
	\item \textit{Short-time transients and late-time non-Markovian dynamics (Section~\ref{sec:diff1}):} 
	The static NHA imposes a constant decay rate from $t=0$, missing the initial Zeno suppression, the gradual buildup of dissipation, and late-time non-Markovian oscillations driven by environmental memory.
	
	\item \textit{Dynamical self-energy reconstruction under quantum quenches (Section~\ref{sec:diff5}):} 
	Treating the environment as a static background misses the continuous evolution and transient relaxation of both the decay rate and the Lamb shift toward their post-quench plateaus.
\end{enumerate}

These examples show that the static NHA provides a useful qualitative reference in the Markovian regime but misses early-time transients, late-time memory effects, and the dynamical response to quenches.
The exact instantaneous self-energy $\Sigma_{\rm inst}(t)$ captures these processes without approximation.
Subsequent papers in this series will examine further limitations associated with strongly energy-dependent self-energies, nonanalytic environmental spectra, and multilevel subsystems.

\section{Summary and Outlook}\label{sec:summary}
The central result of this work is an exact time-local description of the dynamics obtained by projecting global unitary evolution onto a local subsystem.
For an initially unpopulated environment, the projected wavefunction obeys the following equation on every time interval where $U_{\rm SS}(t)$ remains invertible:
\begin{equation}
	i\partial_t |\Psi_{\rm S}(t)\rangle = 
	H_{\rm eff}(t) |\Psi_{\rm S}(t)\rangle,
\end{equation}
with
\begin{equation}
	H_{\rm eff}(t) = i \dot{U}_{\rm SS}(t) U_{\rm SS}^{-1}(t) = H_{\rm S} + \Sigma_{\rm inst}(t).
\end{equation}
This relation follows from an exact algebraic identity and requires no phenomenological assumptions or perturbative approximations.
The instantaneous self-energy $\Sigma_{\rm inst}(t)$ encodes the full environmental backaction, and its anti-Hermitian component quantifies the instantaneous net probability exchange across the subsystem boundary.
Local non-Hermiticity thus arises from restricting the description to a subsystem while the global evolution remains unitary.
This microscopic construction provides a basis for assessing phenomenological non-Hermitian models and their approximations.
The framework has three main features:

\textit{(i) Approximation-free construction:}
The derivation does not invoke the Born or Markov approximation or the wide-band limit.
The construction uses the single-particle direct-sum decomposition $\mathcal{H} = \mathcal{H}_{\rm S} \oplus \mathcal{H}_{\rm B}$, with the time-local generator defined on intervals where $U_{\rm SS}(t)$ is invertible.
For an initially populated environment, the same effective generator appears together with the source term derived in Theorem~II.
Here, non-Hermiticity emerges by projecting the global unitary dynamics onto the subsystem Hilbert space.
This construction differs from reduced-density-matrix approaches, which trace over the environment in a tensor-product Hilbert space, $\mathcal{H}_{\rm S} \otimes \mathcal{H}_{\rm B}$.

\textit{(ii) An exact quantitative benchmark:}
For a static phenomenological Hamiltonian $H_{\rm nH} = H_{\rm S} + \Sigma_0$, the deviation from the exact time-local generator is quantified by $\delta\Sigma(t) = \Sigma_{\rm inst}(t) - \Sigma_0$.
This difference identifies the dynamical processes omitted by the static approximation, including short-time Zeno transients, finite-bandwidth memory, mode-dependent spectral effects, coherent backflow, and finite-size recurrences.
A systematic analysis will follow in \textit{NHQM~II}.

\textit{(iii) A path beyond static non-Hermiticity:} 
The full time-dependent generator $H_{\rm eff}(t) = H_{\rm S} + \Sigma_{\rm inst}(t)$ describes regimes where static approximations become unreliable.
These include short- and long-time dynamics, structured environments, multilevel systems, and nonequilibrium protocols.
Our framework provides an exact microscopic foundation for this time-dependent non-Hermitian description.
Subsequent papers will develop practical extensions of the framework.
In particular, \textit{NHQM~III} will formulate a systematic time-local framework for multilevel systems, including operator-ordering effects.

\appendix
\section*{Appendix}
\makeatletter
\def\appendixtocline#1#2{%
	\vspace{1.5pt}%
	\begingroup
	\parindent \z@ 
	\rightskip \@pnumwidth
	\parfillskip -\@pnumwidth
	\leavevmode
	\advance\leftskip 2.4em\relax
	\hskip -\leftskip
	#1\nobreak\hfil \nobreak\hb@xt@\@pnumwidth{\hss #2}\par
	\endgroup}
\makeatother
\vspace{0.5em}

\appendixtocline{\hyperref[app:Proof1]
{Appendix A: Proof of Theorem I}}{\pageref{app:Proof1}}

\appendixtocline{\hyperref[app:Proof2]
{Appendix B: Proof of Theorem II}}{\pageref{app:Proof2}}

\appendixtocline{\hyperref[app:Proof5]
{Appendix C: Equivalence Between Non-Normality and Noncommutativity}}{\pageref{app:Proof5}}

\appendixtocline{\hyperref[app:FGR]
{Appendix D: Derivation of the On-Shell Self-Energy}}{\pageref{app:FGR}}

\section{Proof of Theorem I}\label{app:Proof1}
In this appendix, we prove Theorem~I and derive the frequency-domain integral representation of the instantaneous self-energy $\Sigma_{\rm inst}(t)$.

\subsubsection{Time-Domain Formulation}
In the direct-sum decomposition $\mathcal{H}=\mathcal{H}_{\rm S}\oplus\mathcal{H}_{\rm B}$, the time-independent global Hamiltonian takes the block form
\begin{equation}
	H = \begin{pmatrix}
		H_{\rm S} & H_{\rm SB} \\[4pt]
		H_{\rm BS} & H_{\rm B}
	\end{pmatrix}.
\end{equation}
The global evolution operator $U(t) = e^{-iHt}$ obeys the corresponding block decomposition:
\begin{equation}
	U(t) = e^{-iHt} =
	\begin{pmatrix}
		U_{\rm SS}(t) & U_{\rm SB}(t) \\[4pt]
		U_{\rm BS}(t) & U_{\rm BB}(t)
	\end{pmatrix}.
\end{equation}
We impose the homogeneous initial condition  $|\Psi_{\rm B}(0)\rangle = 0$.
The global wavefunction $|\Psi(t)\rangle = U(t)|\Psi(0)\rangle$ projected onto the subsystem $\mathcal{H}_{\rm S}$ reads:
\begin{equation}
	|\Psi_{\rm S}(t)\rangle = 
	P_{\rm S}U(t)|\Psi(0)\rangle = 
	U_{\rm SS}(t)|\Psi_{\rm S}(0)\rangle .
	\label{A3}
\end{equation}
We assume that $\mathcal{H}_{\rm S}$ is finite dimensional and restrict the following construction to a time interval on which the projected propagator is invertible:
\begin{equation}
	\det U_{\rm SS}(t)\neq0.
	\label{A4}
\end{equation}
On this interval, differentiating Eq.~\eqref{A3} gives
\begin{equation}
	\begin{aligned}
		i\partial_t \left|\Psi_{\rm S}(t)\right\rangle
		&= i\big[\partial_t U_{\rm SS}(t)\big] \left|\Psi_{\rm S}(0)\right\rangle \\[2pt]
		&= i\big[\partial_t U_{\rm SS}(t)\big] U_{\rm SS}^{-1}(t) \left|\Psi_{\rm S}(t)\right\rangle \\[2pt]
		&= H_{\rm eff}(t) \left|\Psi_{\rm S}(t)\right\rangle.
		\label{A5}
	\end{aligned}
\end{equation}
Decomposing the effective generator as
\begin{equation}
	H_{\rm eff}(t) = H_{\rm S} + \Sigma_{\rm inst}(t), 
	\label{A6}
\end{equation}
defines the instantaneous self-energy,
\begin{equation}
	\Sigma_{\rm inst}(t) = 
	\left[ i\partial_t U_{\rm SS}(t) - H_{\rm S}U_{\rm SS}(t) \right] U_{\rm SS}^{-1}(t).
	\label{A7}
\end{equation}
It converts the time-nonlocal influence of the environment into an exact time-local generator, with all memory effects encoded in the temporal dependence of $\Sigma_{\rm inst}(t)$.

\subsubsection{Frequency-Domain Integral}
We next derive the frequency-domain representation of Eq.~\eqref{A7}.
The inverse Fourier transform of the retarded Green's function gives $-i\Theta(t)U_{\rm SS}(t)$. Thus, for $t>0$, the projected propagator can be written as
\begin{equation}
	U_{\rm SS}(t) = \frac{i}{2\pi} \int_{-\infty}^{\infty} d\omega\, G_{\rm S}^{R}(\omega) e^{-i\omega t},\quad t>0.
	\label{A8}
\end{equation}
Differentiating Eq.~\eqref{A8} with respect to time $t$ gives
\begin{equation}
	i\partial_t U_{\rm SS}(t) = \frac{i}{2\pi} \int_{-\infty}^{\infty} d\omega\, \omega\, G_{\rm S}^{R}(\omega)e^{-i\omega t}.
	\label{A9}
\end{equation}
Subtracting the bare subsystem contribution yields
\begin{equation}
	\begin{aligned}
		&i\partial_t U_{\rm SS}(t) - H_{\rm S}U_{\rm SS}(t) \\[2pt]
		&= \frac{i}{2\pi} \int_{-\infty}^{\infty} d\omega\, \left[\omega I_{\rm S} -H_{\rm S}\right]G_{\rm S}^{R}(\omega)e^{-i\omega t}.
	\end{aligned}
	\label{A10}
\end{equation}
The retarded Green's function in the frequency domain satisfies the exact Dyson-resolvent identity:
\begin{equation}
	\left[(\omega+i0^+)I_{\rm S} - H_{\rm S} - \Sigma_{\rm S}^{R}(\omega)\right]G_{\rm S}^{R}(\omega) = I_{\rm S}.
	\label{A11}
\end{equation}
Rearranging Eq.~\eqref{A11}, we obtain
\begin{equation}
	\left[\omega I_{\rm S} - H_{\rm S}\right]G_{\rm S}^{R}(\omega) = I_{\rm S} + \Sigma_{\rm S}^{R}(\omega)G_{\rm S}^{R}(\omega) - i0^+\,G_{\rm S}^{R}(\omega).
	\label{A12}
\end{equation}
Substituting Eq.~\eqref{A12} into Eq.~\eqref{A10} yields
\begin{equation}
	\begin{aligned}
		i\partial_t U_{\rm SS}(t) - H_{\rm S} U_{\rm SS}(t)
		&= \frac{i}{2\pi}\int_{-\infty}^{\infty} d\omega\, I_{\rm S} e^{-i\omega t} \\[2pt]
		&+ \frac{i}{2\pi}\int_{-\infty}^{\infty}d\omega\, \Sigma_{\rm S}^{R}(\omega) G_{\rm S}^{R}(\omega)e^{-i\omega t}\\[4pt]
		&- i0^+\, U_{\rm SS}(t).
	\end{aligned}
	\label{A13}
\end{equation}
The first term is the contact term $i\delta(t)I_{\rm S}$ and therefore vanishes for fixed $t>0$. The last term vanishes in the retarded limit, with the same positive regulator used in both $\Sigma_{\rm S}^{R}$ and $G_{\rm S}^{R}$ before the limit is taken.
Thus,
\begin{equation}
	\begin{aligned}
		&i\partial_t U_{\rm SS}(t) - H_{\rm S}U_{\rm SS}(t) \\[2pt]
		&= \frac{i}{2\pi} \int_{-\infty}^{\infty} d\omega\, \Sigma_{\rm S}^{R}(\omega) G_{\rm S}^{R}(\omega) e^{-i\omega t}, \quad t>0.
		\label{A14}
	\end{aligned}
\end{equation}
Substituting Eqs.~\eqref{A8} and \eqref{A14} into the exact time-domain identity \eqref{A7} gives
\begin{equation}
	\begin{aligned}
		\Sigma_{\rm inst}(t) &= \left[ \frac{i}{2\pi}\int_{-\infty}^{\infty}d\omega\,\Sigma_{\rm S}^{R}(\omega)G_{\rm S}^{R}(\omega)e^{-i\omega t} \right] \\
		&\quad \times \left[ \frac{i}{2\pi}\int_{-\infty}^{\infty}d\omega\,G_{\rm S}^{R}(\omega)e^{-i\omega t} \right]^{-1}.
	\end{aligned}
	\label{A16}
\end{equation}
This representation applies for $t>0$ on intervals where $U_{\rm SS}(t)$ is invertible; the value at $t=0$ is obtained from the time-domain definition.

\section{Proof of Theorem II}\label{app:Proof2}
In this appendix, we prove Theorem~II on a time interval over which the subsystem propagator $U_{\rm SS}(t)$ is invertible.

\subsubsection{Algebraic Derivation of the Source Term}
Consider a general global initial state with nonzero components in both the subsystem and environment.
The subsystem projection of the global wavefunction is
\begin{equation}
	|\Psi_{\rm S}(t)\rangle = U_{\rm SS}(t)|\Psi_{\rm S}(0)\rangle + U_{\rm SB}(t)|\Psi_{\rm B}(0)\rangle.
	\label{B1}
\end{equation}
Differentiating Eq.~\eqref{B1} with respect to time yields
\begin{equation}
	i\partial_t|\Psi_{\rm S}(t)\rangle = \left[ i\partial_tU_{\rm SS}(t) \right] |\Psi_{\rm S}(0)\rangle + \left[ i\partial_tU_{\rm SB}(t) \right] |\Psi_{\rm B}(0)\rangle.
	\label{B2}
\end{equation}
Because $U_\mathrm{SS}(t)$ is invertible, Eq.~\eqref{B1} can be solved for the initial subsystem component,
\begin{equation}
	|\Psi_{\rm S}(0)\rangle = U_{\rm SS}^{-1}(t) \left[ |\Psi_{\rm S}(t)\rangle - U_{\rm SB}(t)|\Psi_{\rm B}(0)\rangle \right].
	\label{B3}
\end{equation}
Substituting Eq.~\eqref{B3} into the first term of Eq.~\eqref{B2} and using Eq.~\eqref{E10}, we obtain
\begin{equation}
	\begin{aligned}
		&\left[ i\partial_tU_{\rm SS}(t) \right] |\Psi_{\rm S}(0)\rangle \\[2pt]
		&= H_{\rm eff}(t) \left[ |\Psi_{\rm S}(t)\rangle - U_{\rm SB}(t)|\Psi_{\rm B}(0)\rangle \right]\\[2pt]
		&= H_{\rm eff}(t) |\Psi_{\rm S}(t)\rangle - H_{\rm eff}(t) U_{\rm SB}(t)|\Psi_{\rm B}(0)\rangle.
	\end{aligned}
	\label{B4}
\end{equation}
Eq.~\eqref{B2} therefore becomes
\begin{equation}
	\begin{aligned}
		i\partial_t|\Psi_{\rm S}(t)\rangle 
		&= H_{\rm eff}(t)|\Psi_{\rm S}(t)\rangle \\[2pt]
		&+ \left[ i\partial_tU_{\rm SB}(t) - H_{\rm eff}(t)U_{\rm SB}(t) \right] |\Psi_{\rm B}(0)\rangle.
	\end{aligned}
	\label{B5}
\end{equation}
Comparison with the inhomogeneous time-local equation
\begin{equation}
	i\partial_t|\Psi_{\rm S}(t)\rangle = 
	H_{\rm eff}(t)|\Psi_{\rm S}(t)\rangle + |F(t)\rangle,
	\label{B6}
\end{equation}
identifies the source vector as
\begin{equation}
	|F(t)\rangle = \left[ i\partial_tU_{\rm SB}(t) - H_{\rm eff}(t)U_{\rm SB}(t) \right] |\Psi_{\rm B}(0)\rangle.
	\label{B7}
\end{equation}

\subsubsection{Reduction via the Block Equations of Motion}
The global evolution operator satisfies $i\partial_t U(t) = H U(t)$.
The equations of motion for the relevant propagator blocks are therefore
\begin{align}
	i\partial_tU_{\rm SB}(t) &= H_{\rm S}U_{\rm SB}(t) + H_{\rm SB}U_{\rm BB}(t), \label{B8} \\[2pt]
	i\partial_tU_{\rm SS}(t) &= H_{\rm S}U_{\rm SS}(t) + H_{\rm SB}U_{\rm BS}(t). \label{B9}
\end{align}
Substituting Eqs.~\eqref{B8} and \eqref{A6} into Eq.~\eqref{B7}, we find that the bare subsystem contribution cancels, yielding
\begin{equation}
	|F(t)\rangle = 
	\left[ H_{\rm SB}U_{\rm BB}(t) - \Sigma_{\rm inst}(t)U_{\rm SB}(t) \right] |\Psi_{\rm B}(0)\rangle.
	\label{B10}
\end{equation}
Eq.~\eqref{B9}, together with the definition of the instantaneous self-energy, gives
\begin{equation}
	\Sigma_{\rm inst}(t) = H_{\rm SB}U_{\rm BS}(t)U_{\rm SS}^{-1}(t).
	\label{B11}
\end{equation}
Substitution of Eq.~\eqref{B11} into Eq.~\eqref{B10} yields
\begin{equation}
	|F(t)\rangle = H_{\rm SB} \left[ U_{\rm BB}(t) - U_{\rm BS}(t)U_{\rm SS}^{-1}(t)U_{\rm SB}(t) \right] |\Psi_{\rm B}(0)\rangle.
	\label{B12}
\end{equation}
The operator in square brackets is the Schur complement of $U_{\rm SS}(t)$ in the block representation of the global propagator.
We denote it by
\begin{equation}
	S_{\rm B}(t) \equiv U_{\rm BB}(t) - U_{\rm BS}(t) U_{\rm SS}^{-1}(t) U_{\rm SB}(t).
	\label{B13}
\end{equation}

\subsubsection{Reduction of the Schur Complement by Global Unitarity}
Global unitarity $U(t)U^{\dagger}(t)=I$ in block matrix form reads:
\begin{equation}
	\begin{pmatrix}
		U_{\rm SS}(t) & U_{\rm SB}(t) \\[4pt]
		U_{\rm BS}(t) & U_{\rm BB}(t)
	\end{pmatrix}
	\begin{pmatrix}
		U_{\rm SS}^{\dagger}(t) & U_{\rm BS}^{\dagger}(t) \\[4pt]
		U_{\rm SB}^{\dagger}(t) & U_{\rm BB}^{\dagger}(t)
	\end{pmatrix}
	=
	\begin{pmatrix}
		I_{\rm S} & 0 \\[4pt]
		0 & I_{\rm B}
	\end{pmatrix}.
	\label{B14}
\end{equation}
The upper-right and lower-right blocks of Eq.~\eqref{B14} give
\begin{align}
	U_{\rm SS}(t)U_{\rm BS}^{\dagger}(t) + U_{\rm SB}(t)U_{\rm BB}^{\dagger}(t) &= 0, 
	\label{B15} \\[2pt]
	U_{\rm BS}(t)U_{\rm BS}^{\dagger}(t) + U_{\rm BB}(t)U_{\rm BB}^{\dagger}(t) &= I_{\rm B}.
	\label{B16}
\end{align}
Since $U_{\rm SS}(t)$ is invertible, Eq.~\eqref{B15} implies
\begin{equation}
	U_{\rm BS}^\dag(t) = -U_{\rm SS}^{-1}(t)U_{\rm SB}(t)U_{\rm BB}^\dag(t).
	\label{B17}
\end{equation}
Substitution into Eq.~\eqref{B16} gives
\begin{equation}
	\left[U_{\rm BB}(t) - U_{\rm BS}(t)U_{\rm SS}^{-1}(t)U_{\rm SB}(t)\right] 
	U_{\rm BB}^\dag(t) = I_{\rm B}.
\end{equation}
In terms of the Schur complement,
\begin{equation}
	S_{\rm B}(t) U_{\rm BB}^\dagger(t) = I_{\rm B}.
	\label{B19}
\end{equation}

To establish that $S_{\rm B}(t)$ is the two-sided inverse of $U_{\rm BB}^\dagger(t)$, we must use $U^\dagger(t)U(t)=I$.
Its lower-left and lower-right blocks give
\begin{align}
	U_{\rm SB}^\dagger(t) U_{\rm SS}(t) + U_{\rm BB}^\dagger(t) U_{\rm BS}(t) &= 0,
	\label{B20} \\[2pt]
	U_{\rm SB}^\dagger(t) U_{\rm SB}(t) + U_{\rm BB}^\dagger(t) U_{\rm BB}(t) &= I_{\rm B}.
	\label{B21}
\end{align}
Right-multiplying Eq.~\eqref{B20} by $U_{\rm SS}^{-1}(t)U_{\rm SB}(t)$ gives
\begin{equation}
	U_{\rm BB}^\dagger(t) U_{\rm BS}(t) U_{\rm SS}^{-1}(t) U_{\rm SB}(t) = -U_{\rm SB}^\dagger(t) U_{\rm SB}(t).
	\label{B22}
\end{equation}
Using Eqs.~\eqref{B21} and \eqref{B22}, we find
\begin{equation}
		U_{\rm BB}^{\dagger}(t) S_{\rm B}(t) = I_{\rm B}.
		\label{B23}
\end{equation}
Equations~\eqref{B19} and \eqref{B23} show that $S_{\rm B}(t)$ is the two-sided inverse of $U_{\rm BB}^\dagger(t)$,
\begin{equation}
	S_{\rm B}(t) = \left[ U_{\rm BB}^\dagger(t) \right]^{-1}.
	\label{B24}
\end{equation}
Substitution into Eq.~\eqref{B12} gives the compact expression
\begin{equation}
	|F(t)\rangle = H_{\rm SB}S_{\rm B}(t)|\Psi_{\rm B}(0)\rangle.
	\label{B25}
\end{equation}

\subsubsection{Interpretation via the Variation-of-Constants Formula}

For the following integral representation with initial time $0$, we assume that $U_{\rm SS}(\tau)$ remains invertible for every $\tau\in[0,t]$.
The formal solution of the inhomogeneous equation
\begin{equation}
	i\partial_t|\Psi_{\rm S}(t)\rangle = H_{\rm eff}(t)|\Psi_{\rm S}(t)\rangle + |F(t)\rangle
\end{equation}
is given by the variation-of-constants formula
\begin{equation}
	|\Psi_{\rm S}(t)\rangle = U_{\rm eff}(t,0)|\Psi_{\rm S}(0)\rangle - i\int_0^t dt'\, U_{\rm eff}(t,t')|F(t')\rangle.
	\label{B27}
\end{equation}
Here, $U_{\rm eff}(t,t')$ is the propagator generated by the time-local effective Hamiltonian,
\begin{equation}
	U_{\rm eff}(t,t') \equiv \mathcal{T} \exp\left[-i\int_{t'}^{t}d\tau\, H_{\rm eff}(\tau)\right].
\end{equation}
On an interval where $U_{\rm SS}(t)$ remains invertible, this propagator satisfies
\begin{equation}
	U_{\rm eff}(t,t') = U_{\rm SS}(t) U^{-1}_{\rm SS}(t').
\end{equation}
In particular, because $U_{\rm SS}(0)=I_{\rm S}$,
\begin{equation}
	U_{\rm eff}(t,0) = U_{\rm SS}(t).
\end{equation}
Using Eq.~\eqref{B7}, one finds
\begin{equation}
	U_{\rm SS}^{-1}(t') \left| F(t') \right\rangle = 
	i \partial_{t'}\! \left[ U_{\rm SS}^{-1}(t') U_{\rm SB}(t') \right] \left| \Psi_{\rm B}(0) \right\rangle.
	\label{B31}
\end{equation}
Consequently, the source contribution in Eq.~\eqref{B27} becomes
\begin{equation}
	\begin{aligned}
		&- i\int_{0}^{t}dt'\,U_{\text{eff}}(t,t')\left|F(t')\right\rangle \\
		&= -iU_{\rm SS}(t) \int_{0}^{t}dt'\,U_{\rm SS}^{-1}(t')\left|F(t')\right\rangle \\[2pt]
		&= U_{\rm SS}(t)\left[U_{\rm SS}^{-1}(t)U_{\rm SB}(t)-U_{\rm SS}^{-1}(0)U_{\rm SB}(0)\right]\left|\Psi_{\rm B}(0)\right\rangle.
	\end{aligned}
	\label{B32}
\end{equation}
Since $U_{\rm SS}(0)=I_{\rm S}$ and $U_{\rm SB}(0)=0$,
Eq.~\eqref{B32} reduces to
\begin{equation}
	-i\int_{0}^{t}dt'U_{\text{eff}}(t,t') \left|F(t')\right\rangle =
	U_{\rm SB}(t)|\Psi_{\rm B}(0)\rangle.
	\label{B33}
\end{equation}
Substitution into Eq.~\eqref{B27} exactly reproduces the projected block evolution,
\begin{equation}
	\left| \Psi_{\rm S}(t) \right\rangle = 
	U_{\rm SS}(t) \left| \Psi_{\rm S}(0) \right\rangle + 
	U_{\rm SB}(t) \left| \Psi_{\rm B}(0) \right\rangle.
	\label{B34}
\end{equation}
Thus, the source term reconstructs the component of the subsystem amplitude that originates from the initial environmental wavefunction and then propagates into the subsystem.
On an invertibility interval that does not contain $0$, the variation-of-constants formula must instead use an initial time $t_0$ within that interval, the state $|\Psi_{\rm S}(t_0)\rangle$, and the lower integration limit $t_0$.

\section{Equivalence Between Non-Normality and Noncommutativity}\label{app:Proof5}
In this appendix, we establish the equivalence between the non-normality of the effective Hamiltonian and the noncommutativity of its Hermitian and anti-Hermitian components.

Let $A(t) \equiv H_{\rm S} + \Delta(t)$ denote the Hermitian part of the effective Hamiltonian.  
Then the effective Hamiltonian and its adjoint can be expressed as
\begin{equation}
	H_{\rm eff}(t) = A(t) - i\frac{\Gamma(t)}{2},\,\,\, H_{\rm eff}^\dagger(t) = A(t) + i\frac{\Gamma(t)}{2}.
\end{equation}
The commutator of the effective Hamiltonian with its adjoint reads
\begin{equation}
	\begin{aligned}
		[H_{\rm eff}(t), H_{\rm eff}^\dagger(t)] &=
		\left[ A(t) - i\frac{\Gamma(t)}{2}, A(t) + i\frac{\Gamma(t)}{2} \right] \\
		&= \frac{i}{2}[A(t),\Gamma(t)] - \frac{i}{2}[\Gamma(t),A(t)] \\[2pt]
		&= i[A(t), \Gamma(t)] \\[2pt]
		&= i[H_{\rm S} + \Delta(t), \Gamma(t)].
	\end{aligned}
\end{equation}
Consequently, $H_{\rm eff}(t)$ is normal if and only if its Hermitian component commutes with its anti-Hermitian component.
Equivalently, the effective Hamiltonian is non-normal precisely when
\begin{equation}
	[H_{\rm S} + \Delta(t), \Gamma(t)] \neq 0.
\end{equation}

\section{Derivation of the On-Shell Self-Energy}\label{app:FGR}
In this appendix, we derive the on-shell self-energy and the associated FGR decay rate for the Anderson–Fano model.

\subsubsection{The Anderson-Fano Model and Impurity Green’s Function}
The global single-particle Hamiltonian reads:
\begin{equation}
	\begin{aligned}
		H &= V_0 |0\rangle\langle 0|  + \lambda \left( |0\rangle\langle 1| + |1\rangle\langle 0| \right) \\
		&\quad + t_c \sum_{j=1}^{N-1} \left( |j\rangle\langle j+1| + |j+1\rangle\langle j| \right),
	\end{aligned}
\end{equation}
where $|0\rangle$ denotes the impurity state and $|j\rangle$, with $j=1,\ldots,N$, denotes the state localized at the $j$th bath site. We take $V_0$ and $\lambda$ to be real and $t_c>0$.
The impurity is coupled to the boundary site of the chain with strength $\lambda$.
The retarded Green’s function of the impurity is
\begin{equation}
	G_{\rm S}^R(\omega) = \frac{1}{\omega + i0^+ - V_0 - \Sigma_{\rm S}^R(\omega)}.
\end{equation}
For a finite bath, the retarded self-energy is the discrete spectral sum
\begin{equation}
	\Sigma_{\rm S}^R(\omega) = \lambda^2 \sum_{m=1}^{N} \frac{|\langle 1 | k_m \rangle|^2}{\omega - \epsilon_{k_m} + i0^+}
	\label{D3}
\end{equation}
In the thermodynamic limit $N\to\infty$, taken before the retarded regulator is sent to zero, we obtain
\begin{equation}
	\Sigma_{\rm S}^R(\omega) = \lambda^2 \int_{-2t_c}^{2t_c} \frac{\rho_1(\epsilon)}{\omega - \epsilon + i0^+}\, d\epsilon,
	\label{D4}
\end{equation}
where $\rho_1(\epsilon)$ is the local density of states (LDOS) at the boundary site of the uncoupled bath.

\subsubsection{Boundary LDOS of the Semi-Infinite 1D Chain}\label{app:LDOS}
For a finite chain comprising $N$ sites with open boundaries, the eigenvalues and normalized eigenstates are
\begin{equation}
		\epsilon_{k_m} = 2t_c \cos\left(\frac{m\pi}{N+1}\right), \quad  m = 1, 2, \ldots, N,
\end{equation}
and
\begin{equation}
	|k_m\rangle = \sqrt{\frac{2}{N+1}} \sum_{j=1}^{N} \sin\left(\frac{m\pi j}{N+1}\right) |j\rangle.
\end{equation}
The spectral weight of mode $m$ at the boundary site $j=1$ is
\begin{equation}
	|\langle 1|k_m\rangle|^2 = \frac{2}{N+1} \sin^2\left(\frac{m\pi}{N+1}\right).
\end{equation}
The corresponding finite-size boundary LDOS is therefore
\begin{equation}
	\rho_{1,N}(\epsilon) = \sum_{m=1}^{N}
	|\langle 1|k_m\rangle|^2 \delta(\epsilon - \epsilon_{k_m}).
\end{equation}
Using
\begin{equation}
	\sum_{m=1}^{N} \to \frac{N+1}{\pi} \int_{0}^{\pi} dk,
\end{equation}
we obtain
\begin{equation}
	\rho_1(\epsilon) = \frac{2}{\pi} \int_{0}^{\pi} dk \sin^2 k \,\delta\left(\epsilon - 2t_c \cos k\right).
\end{equation}
For $t_c>0$ and $|\epsilon|<2t_c$, the equation $\epsilon=2t_c \cos k$ has one root $k_{\epsilon}\in(0,\pi)$, with
\begin{equation}
	\sin k_{\epsilon} = \frac{\sqrt{4t_c^2-\epsilon^2}}{2t_c}.
\end{equation}
Using the delta-function transformation
$\delta\left(\epsilon - 2t_c \cos k\right) = \delta(k-k_\epsilon)/(2t_c\sin k_\epsilon)$,
the boundary LDOS becomes
\begin{equation}
	\rho_1(\epsilon) = \frac{1}{2\pi t_c^2} 
	\sqrt{4t_c^2 - \epsilon^2} 
	\,\Theta(2t_c - |\epsilon|).
	\label{D12}
\end{equation}
Unlike the bulk DOS of an infinite 1D chain, the boundary LDOS vanishes at the band edges.

\subsubsection{Retarded Self-Energy}
Using the Sokhotski–Plemelj identity,
\begin{equation}
	\frac{1}{\omega - \epsilon + i0^+} = 
	\mathcal{P} \frac{1}{\omega - \epsilon} -
	i\pi\delta(\omega - \epsilon),
\end{equation}
the self-energy [Eq.~\eqref{D4}] can be decomposed as
\begin{equation}
	\Sigma_{\rm S}^R(\omega) = \Delta(\omega) - \frac{i}{2} \Gamma(\omega),
\end{equation}
where
\begin{equation}
	\Gamma(\omega) = 2\pi\lambda^2 \rho_1(\omega) = \frac{\lambda^2}{t_c^2}\sqrt{4t_c^2-\omega^2}
	\,\Theta(2t_c - |\omega|),
	\label{D15}
\end{equation}
and
\begin{equation}
	\Delta(\omega) = \lambda^2 \mathcal{P}\int_{-2t_c}^{2t_c} \frac{\rho_1(\epsilon)}{\omega - \epsilon} d\epsilon.
	\label{D16}
\end{equation}
For $|\omega|<2t_c$, the principal-value integral evaluates to
\begin{equation}
	\Delta(\omega) = \frac{\lambda^2}{2t_c^2}\omega.
\end{equation}
Therefore, the retarded self-energy takes the form:
\begin{equation}
	\Sigma_{\rm S}^R(\omega) = \frac{\lambda^2}{2t_c^2} \left( \omega - i\sqrt{4t_c^2 - \omega^2} \right), \quad |\omega|<2t_c.
	\label{D18}
\end{equation}

\subsubsection{On-Shell Approximation and Time-Domain Dynamics}
For an impurity energy inside the bath band, $|V_0|<2t_c$, the on-shell approximation replaces $\Sigma_{\rm S}^R(\omega)$ by its value at $\omega=V_0$:
\begin{equation}
	\begin{aligned}
		\Sigma_{\rm S}^R(V_0)  &= \frac{\lambda^2}{2t_c^2} \left( V_0 - i\sqrt{4t_c^2 - V_0^2} \right) \\
		&= \Delta_{\rm FGR} - i\frac{\Gamma_{\rm FGR}}{2}.
	\end{aligned}
\end{equation}
The corresponding decay rate is 
\begin{equation}
	\Gamma_{\rm FGR} = 
	\frac{\lambda^2}{t_c^2} \sqrt{4t_c^2 - V_0^2}.
\end{equation}
With the replacement $\Sigma_{\rm S}^R(\omega)\approx\Sigma_{\rm S}^R(V_0)$, we obtain the static non-Hermitian Hamiltonian
\begin{equation}
	H_{\rm nH} = V_0 + \Delta_{\rm FGR} - i\Gamma_{\rm FGR}/2.
\end{equation}
For an initial state localized on the impurity, inverse Fourier transformation of the approximate Green's function for $t>0$ gives the survival probability
\begin{equation}
	P_0^{\rm NHA}(t) = e^{-\Gamma_{\rm FGR}t},
\end{equation}
which is purely exponential within the static approximation. This result does not describe exact finite-band dynamics at all times or the recurrences of a finite bath.

\makeatletter
\addtocontents{toc}{\protect\let\protect\l@section\protect\oldl@section}
\addcontentsline{toc}{section}{\protect\numberline{VI}~References}
\let\oldaddcontentsline\addcontentsline
\renewcommand{\addcontentsline}[3]{}
\makeatother
\bibliography{references}
\makeatletter
\let\addcontentsline\oldaddcontentsline
\makeatother
\end{document}